\pdfoutput=1 
\documentclass{article}

\usepackage{spconf}
\usepackage{amsmath,amssymb}
\usepackage{graphicx}
\usepackage{times}
\usepackage{latexsym}
\usepackage[T1]{fontenc}     
\usepackage[utf8]{inputenc}  
\usepackage{microtype}       
\usepackage{inconsolata}     
\usepackage{xcolor}
\PassOptionsToPackage{breaklinks}{hyperref}
\usepackage{hyperref}
\definecolor{darkblue}{rgb}{0, 0, 0.5}
\hypersetup{colorlinks=true, citecolor=darkblue, linkcolor=darkblue, urlcolor=darkblue}

\usepackage{booktabs}
\usepackage{array}      
\usepackage{multirow}   
\usepackage{diagbox}    
\usepackage{pifont}     
\usepackage{wasysym}    

\usepackage{algorithm}
\usepackage{algpseudocode}

\makeatletter
\renewcommand\paragraph[1]{\par\vspace{1.5ex plus 0.5ex minus .2ex}\noindent{\normalsize\bfseries #1}\hspace{1em}\ignorespaces}
\makeatother
\let\origtexttt\texttt
\renewcommand{\texttt}[1]{\origtexttt{\hyphenchar\font=`\-\relax #1}}

\newcommand{\benchname}{TORUS}
\newcommand{\pms}[1]{{\scriptsize$\pm$#1}}  

\definecolor{tickgreen}{RGB}{27,138,90}
\definecolor{crossred}{RGB}{198,58,58}
\newcommand{\gmark}{\textcolor{tickgreen}{\ding{51}}}     
\newcommand{\rmark}{\textcolor{crossred}{\ding{55}}}      
\newcommand{\pmark}{\textcolor{black!55}{$\bullet$}}      

\newcommand{\lnvidia}{\raisebox{-0.22ex}{\includegraphics[height=1.9ex]{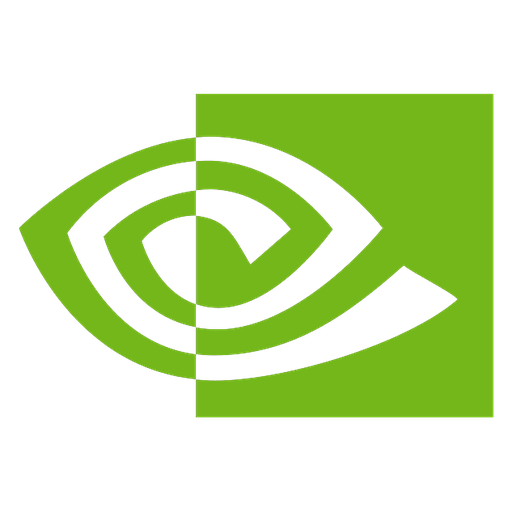}}\hspace{0.25em}}           
\newcommand{\ltencent}{\raisebox{-0.22ex}{\includegraphics[height=1.9ex]{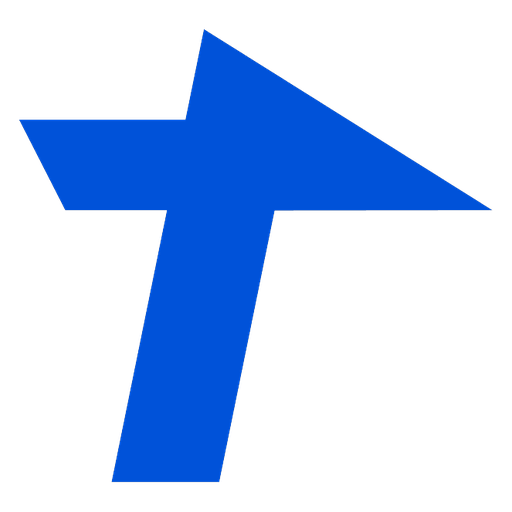}}\hspace{0.25em}}         
\newcommand{\lai}{\raisebox{-0.22ex}{\includegraphics[height=1.9ex]{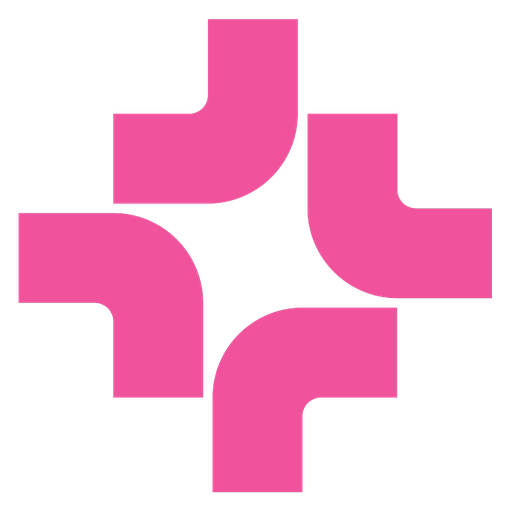}}\hspace{0.25em}}                  
\newcommand{\lopenai}{\raisebox{-0.22ex}{\includegraphics[height=1.9ex]{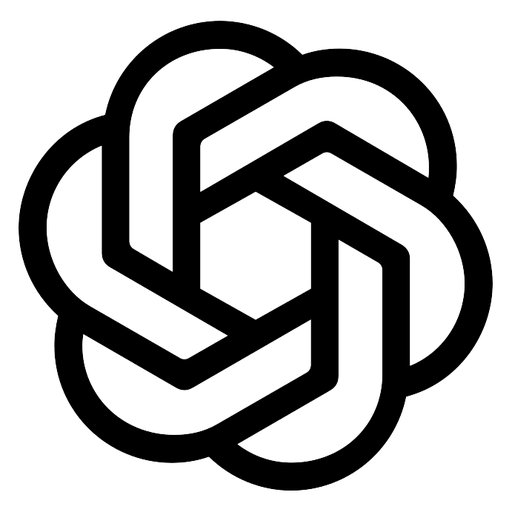}}\hspace{0.25em}}           
\newcommand{\lgemini}{\raisebox{-0.22ex}{\includegraphics[height=1.9ex]{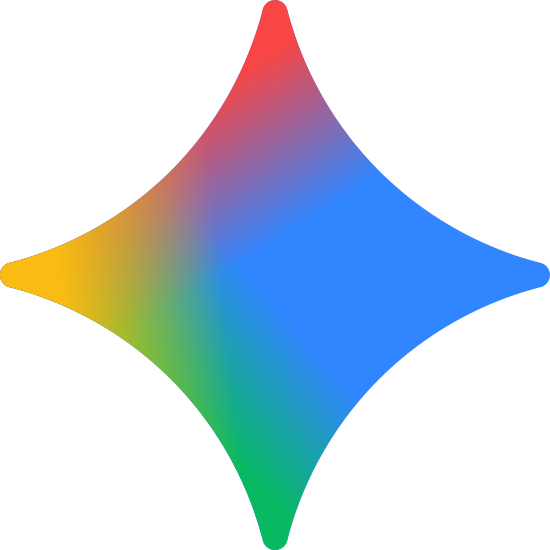}}\hspace{0.25em}}
\newcommand{\lnone}{}  
\newcommand{\gptaudio}{\lopenai\texttt{gpt-audio}}
\newcommand{\toruslogo}{\raisebox{-18pt}{\includegraphics[height=40pt]{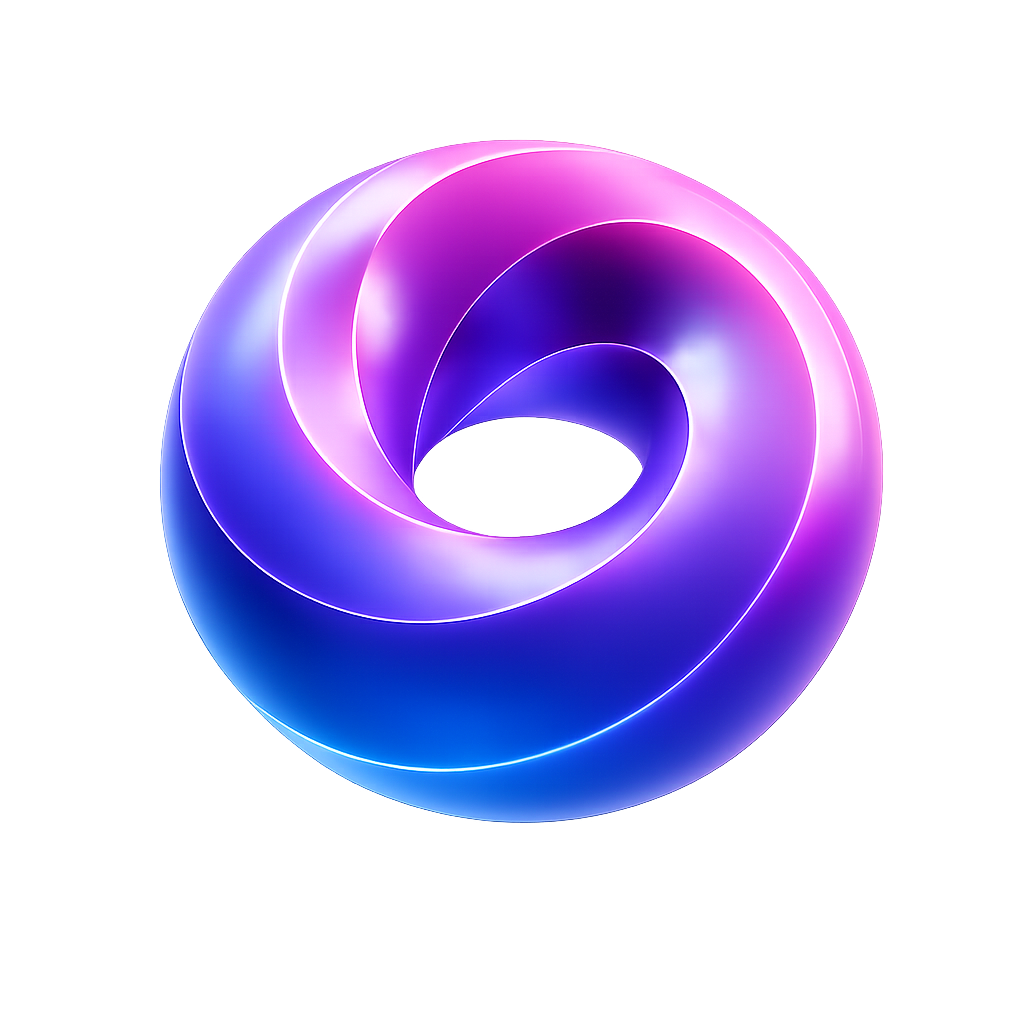}}}

\newcommand{\lself}{\raisebox{-0.2ex}{\includegraphics[height=1.8ex]{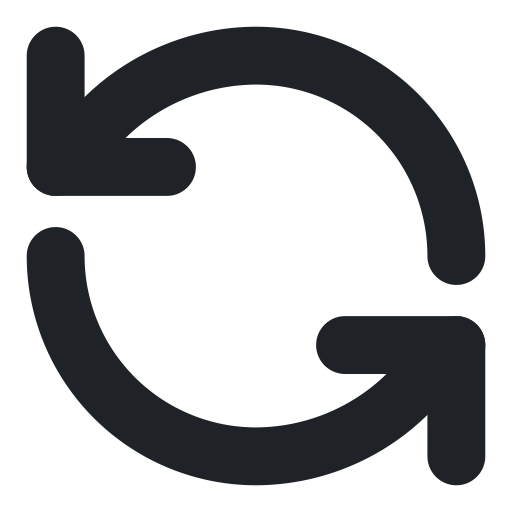}}\hspace{0.15em}}    
\newcommand{\lcasc}{\raisebox{-0.2ex}{\includegraphics[height=1.8ex]{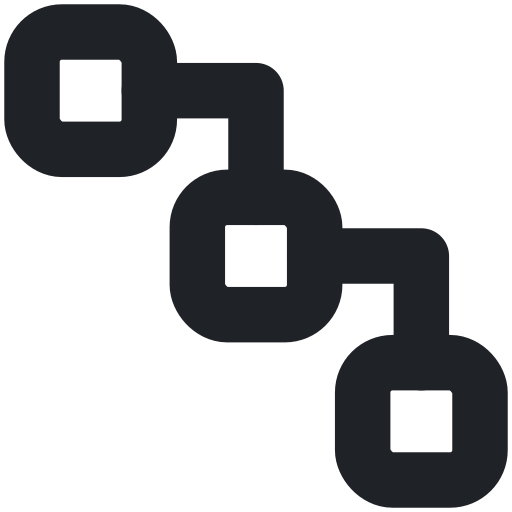}}\hspace{0.3em}}     
\newcommand{\lnative}{\raisebox{-0.2ex}{\includegraphics[height=1.8ex]{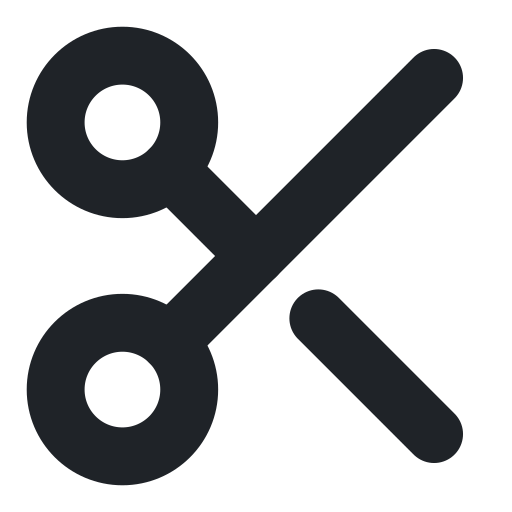}}}               
\newcommand{\lselfcap}{\raisebox{-0.2ex}{\includegraphics[height=1.8ex]{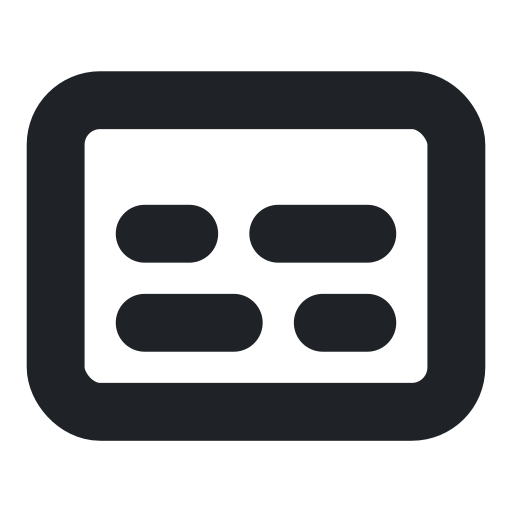}}}             
\newcommand{\lta}{\raisebox{-0.2ex}{\includegraphics[height=1.8ex]{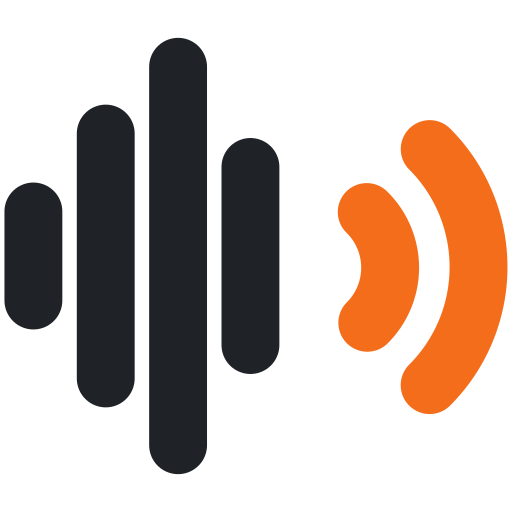}}}                      
\newcommand{\ltts}{\raisebox{-0.2ex}{\includegraphics[height=1.8ex]{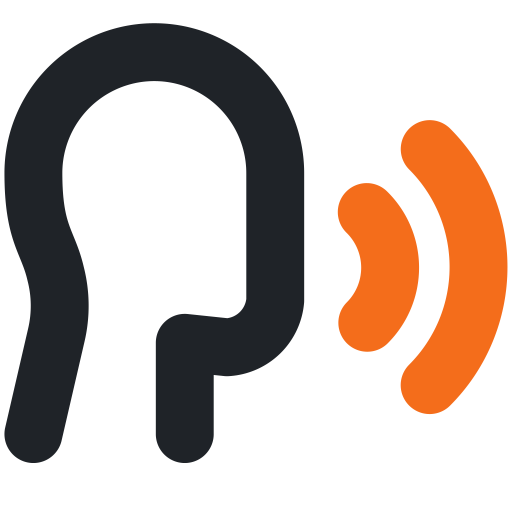}}}                     
\newcommand{\lsp}{\raisebox{-0.2ex}{\includegraphics[height=1.8ex]{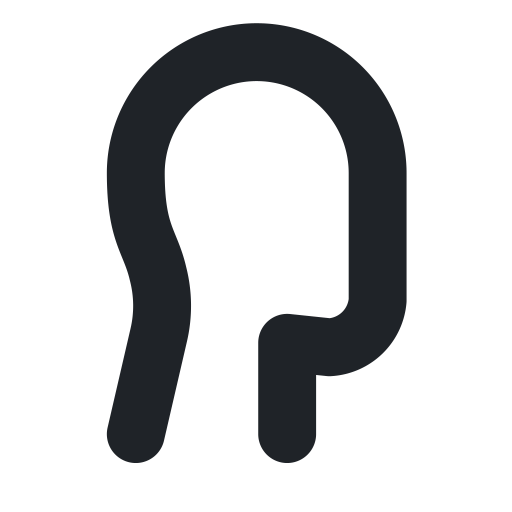}}}                   
\newcommand{\lso}{\raisebox{-0.2ex}{\includegraphics[height=1.8ex]{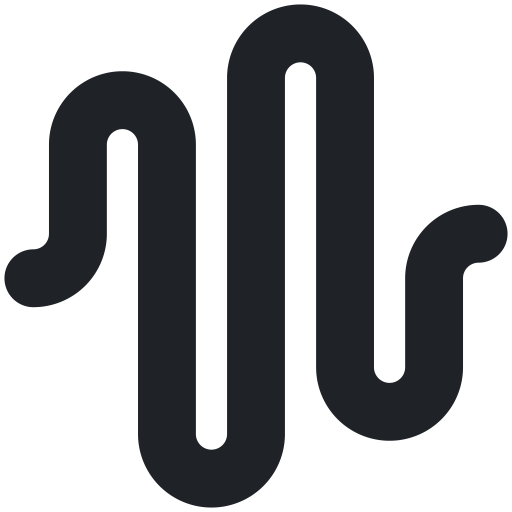}}}                    
\newcommand{\lmu}{\raisebox{-0.2ex}{\includegraphics[height=1.8ex]{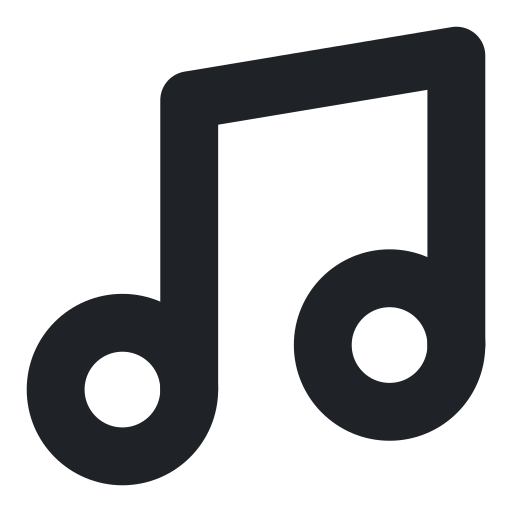}}}                    

\newcommand{\ugco}{UGC$_{\mathrm{OWN}}$}  
\newcommand{\ugcf}{UGC$_{\mathrm{F}}$}    

\title{~\toruslogo~Unifying Benchmarking for Unified Audio Models \vspace{-15pt}}
\name{Aryan Vijay Bhosale$^{1,2}$, Harshit Rajgarhia$^{1}$, Abhishek Mukherji$^{1\dagger}$, Dinesh Manocha$^{2\dagger}$}
\address{
  $^{1}$Centific Global Solutions Inc. \quad
  $^{2}$University of Maryland, College Park \vspace{5pt}\\  \texttt{aryan.vi.bhosale@gmail.com, harshit.rajgarhia@centific.com}
}

\begin{document}
\maketitle
{\renewcommand{\thefootnote}{}%
\footnotetext{\hspace*{-1.8em} $^{\dagger}$Equal advising. Work done during an internship at Centific AI Research. \\ \hspace*{0.4em} \texttt{aryan.vi.bhosale@gmail.com, harshit.rajgarhia@centific.com}}

\begin{abstract}
Unified audio models capable of audio understanding, audio generation and, increasingly, audio editing are proliferating rapidly. However, each capability is evaluated in isolation on specialized benchmarks and a basic question remains unanswered: can these models understand the audio they generate/edit i.e. do they exhibit audio understanding-generation self-coherence (\ugco)? We introduce \benchname{}, an audio-native unified benchmark for holistic, closed-loop evaluation of unified audio models comprising 1{,}053 questions (117 tests) spanning 21 tasks and 8 audio domain settings across all 3 audio types. We evaluate unified models and cascades of specialized models across objective and subjective metrics. Best unified models are able to answer only 52.4\% questions about their generations, while \lopenai\texttt{gpt-audio} can answer 53.6\% questions. We hypothesize that a truly unified audio intelligence should be highly self-coherent and thus position understanding-generation coherence as an essential test for future unified audio systems. Further details are on our \href{https://torus-benchmark.github.io/}{project page}.
\end{abstract}

\section{Introduction}
\label{sec:intro}

On the quest for audio general intelligence, we see the emergence of  unified audio models~\cite{uniaudio2_2026,tian2026audioomni,lu2024unifiedio2} capable of audio generation, understanding and increasingly editing. They echo the belief that representations shared across tasks let each capability inform the others. Most such models have two heads: a \emph{generation head} $G$ that generates/edits audio and an audio \emph{understanding head} $U$. Such models report numbers on specialized benchmarks: audio question-answering benchmarks~\cite{mmaupro2025, sakshi2025mmau}, generation on fidelity and adherence suites~\cite{ttabench2025,ritta2025} and editing on instruction-following metrics~\cite{speecheditbench2026,mmae2026}. A model can place well on all three leaderboards while nothing
certifies that its two heads agree about the same
piece of audio (Fig.~\ref{fig:motivation}). No benchmark measures the \emph{unified} property these models actually claim. Unified benchmarks exist for vision models~\cite{umnibench2025,mmeunify2025}, but their protocol does not transfer directly since audio is inherently temporal and questions can carry unusually strong text priors~\cite{glitters2026audio}. An audio-native unified coherence benchmark must therefore be constructed against text-prior leakage and around audio’s own tasks transcending tasks inspired by vision. To tackle these issues, we construct the \benchname{} (\textbf{T}est \textbf{o}f \textbf{R}endering-\textbf{U}nderstanding \textbf{S}elf-Coherence) Benchmark comprising 117 tests across sound, speech and music. We quantify understanding-generation coherence (UGC) in terms of \% accuracy on 1{,}053 six-option multiple choice questions (chance 16.7\%). We define self-coherence (\ugco) as a measure of the models ability to understand its own generations and discuss how this depends on multiple factors like generation fidelity, strength of comprehension, etc. To attribute performance of both heads and identify bottlenecks, we construct \ugcf by using a frontier understanding model (to measure coherence with a stronger understanding head i.e. fewer understanding failures) instead of the models own understanding head and report audio generation metrics WER, KL, FAD, CM, PQ (total 7 objective metrics) and perform human evaluation.

\begin{figure}[t]
    \centering
    \includegraphics[width=0.85\linewidth]{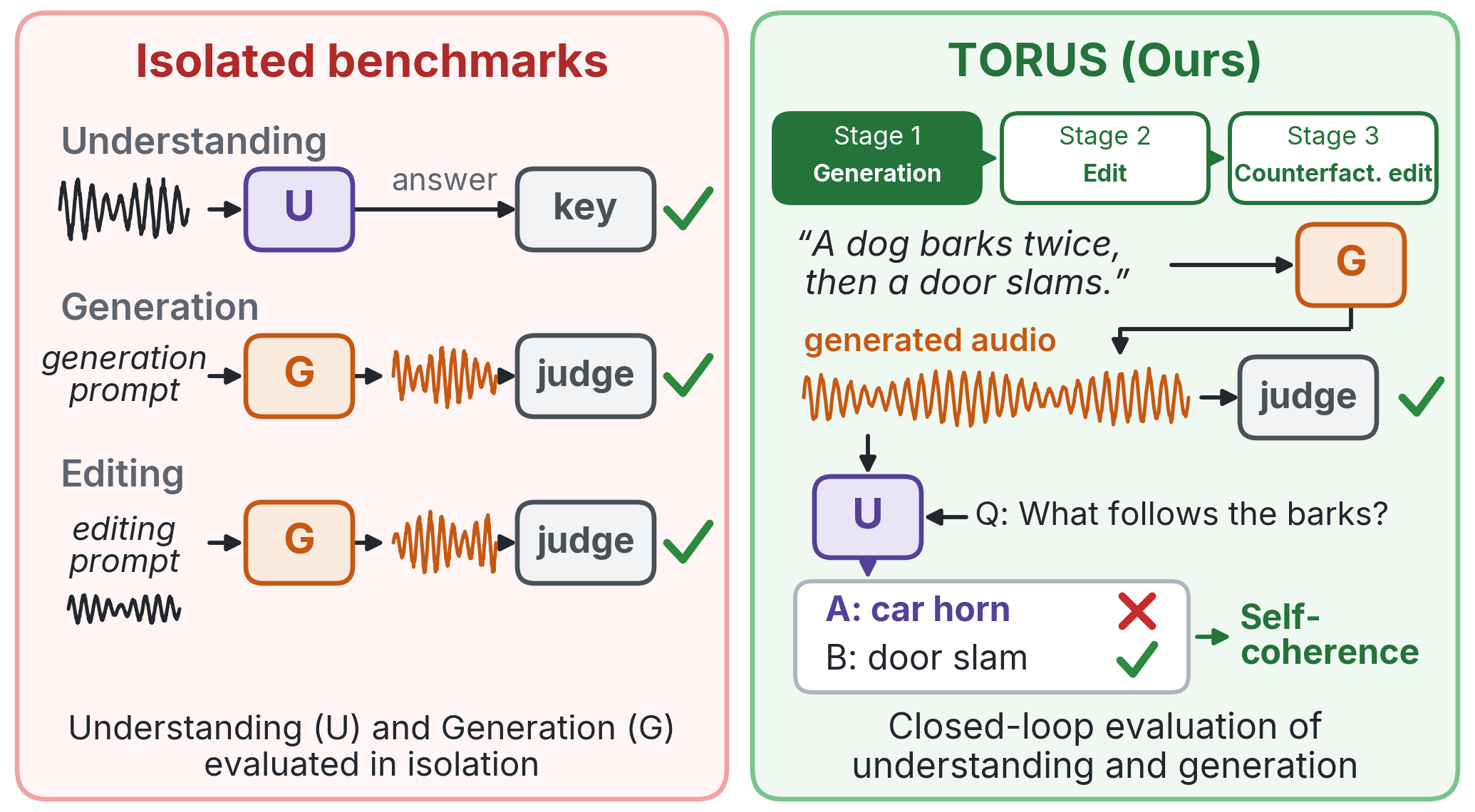}
    \caption{Comparison of \textbf{task-specific} (specialized audio understanding, generation and editing) v/s \textbf{closed-loop evaluation} presented through an illustrative example.}
    \label{fig:motivation}
\end{figure}

\begin{figure*}[t]
    \centering
    \includegraphics[width=\textwidth]{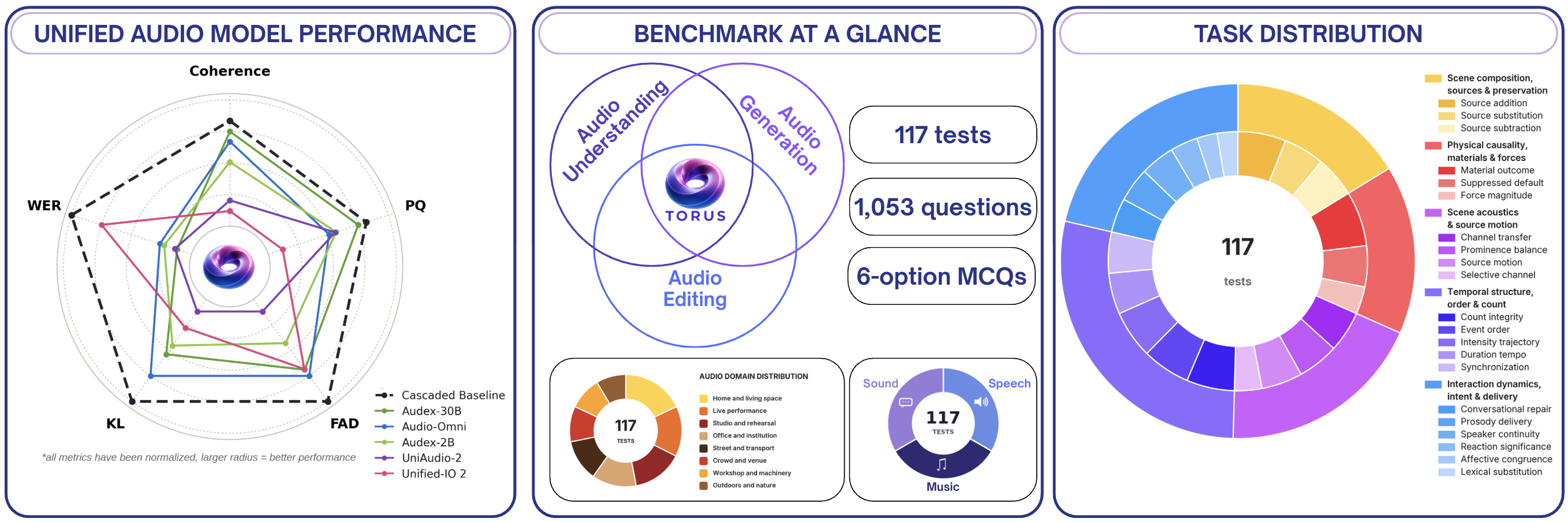}
    \vspace{-10pt}
    \caption{Left: \textbf{Performance of audio models} on 5 metrics (normalized s.t. larger is better). Center: \benchname{} contains \textbf{117} tests i.e. \textbf{1{,}053} six-option MCQs. Audio type (\textbf{3}) \& domain (\textbf{8}) distribution. Right: Benchmark task distribution: \textbf{21} tasks, \textbf{5} task families.}
    \label{fig:spider_and_stats}
\end{figure*}

\noindent\textbf{Main Contribution.} In this work, we release \benchname{}, the \textbf{first audio-native benchmark} for unified audio models. We present the recipe for a human-verified, text-prior leak-checked test generation pipeline and curate \textbf{117} tests to probe understanding-generation coherence. We benchmark 5 open unified models and a baseline of cascaded specialized models on \textbf{1{,}053 MCQs} along 8 axes. We find that evaluated baselines exhibit \textbf{limited understanding-generation coherence} with best models scoring 52.4\% self-coherence (and 53.6\% \ugcf) and hope to motivate research on unified audio models that are self-coherent.



\section{Related Work}
\label{sec:related}
\textbf{Specialized Audio Benchmarks.} Understanding benchmarks
(MMAU-Pro~\cite{mmaupro2025}, ADQA-Bench~\cite{he2026dcase}) score text answers
about \emph{given} audio, while generation (TTA-Bench~\cite{ttabench2025},
RiTTA~\cite{ritta2025}) and editing (SpeechEditBench~\cite{speecheditbench2026},
MMAE~\cite{mmae2026}) benchmarks score the audio a model produces with
objective metrics and external judges. Each evaluates one capability, and none
asks a model about audio it produced (Table~\ref{tab:benchcompare}).\\
\noindent\textbf{Unified Multimodal Benchmarking.} Unified audio models
(UniAudio-2~\cite{uniaudio2_2026}, Audex~\cite{kong2026audex},
Unified-IO~2~\cite{lu2024unifiedio2}, Audio-Omni~\cite{tian2026audioomni})
generate, understand and, in some cases, edit audio in one system. In vision,
MME-Unify~\cite{mmeunify2025} and UmniBench~\cite{umnibench2025} question a model about images it generated, across dependent stages. In audio, AQAScore~\cite{kuan2026aqascore} questions generated clips, but
a separate model answers.

\begin{table}[t]
\centering
\footnotesize
\setlength{\tabcolsep}{3pt}
\renewcommand{\arraystretch}{1.05}
\begin{tabular}{@{}l r cccc ccc cc@{}}
\toprule
& & \multicolumn{4}{c}{\textbf{Modality}} & \multicolumn{3}{c}{\textbf{Tasks}} & \multicolumn{2}{c}{\textbf{Protocol}} \\
\cmidrule(lr){3-6} \cmidrule(lr){7-9} \cmidrule(l){10-11}
\textbf{Benchmark} & \textbf{Size} & V & Sp & So & Mu & U & G & E &
\textbf{CF} & \textbf{Leak} \\
\midrule
MMAU-Pro~\cite{mmaupro2025}                & 5{,}305  & \rmark & \gmark & \gmark & \gmark & \gmark & \rmark & \rmark & \rmark & \pmark \\
ADQA-Bench~\cite{he2026dcase}                 & 3{,}000  & \rmark & \gmark & \gmark & \gmark & \gmark & \rmark & \rmark & \rmark & \gmark \\
RiTTA~\cite{ritta2025}                     & 800      & \rmark & \rmark & \gmark & \rmark & \rmark & \gmark & \rmark & \rmark & \rmark \\
TTA-Bench~\cite{ttabench2025}              & 2{,}999  & \rmark & \gmark & \gmark & \gmark & \rmark & \gmark & \rmark & \rmark & \rmark \\
SpeechEditB.~\cite{speecheditbench2026}    & 4{,}700  & \rmark & \gmark & \rmark & \rmark & \rmark & \rmark & \gmark & \rmark & \rmark \\
MMAE~\cite{mmae2026}                       & 2{,}000  & \rmark & \gmark & \gmark & \gmark & \rmark & \rmark & \gmark & \rmark & \rmark \\
MME-Unify~\cite{mmeunify2025}              & 4{,}104  & \gmark & \rmark & \rmark & \rmark & \gmark & \gmark & \gmark & \rmark & \rmark \\
UmniBench~\cite{umnibench2025}             & 4{,}797  & \gmark & \rmark & \rmark & \rmark & \gmark & \gmark & \gmark & \gmark & \rmark \\
\midrule
\textbf{\benchname{} (ours)} & 1{,}053 & \rmark & \gmark & \gmark & \gmark & \gmark & \gmark & \gmark & \gmark & \gmark \\
\bottomrule
\end{tabular}
\caption{\textbf{Comparison with existing benchmarks} along size, audio type, tasks, counterfactual testing and audio/text question/prompt leakage screening.}
\label{tab:benchcompare}
\end{table}
\begin{figure*}[t]
\centering
\includegraphics[width=\linewidth]{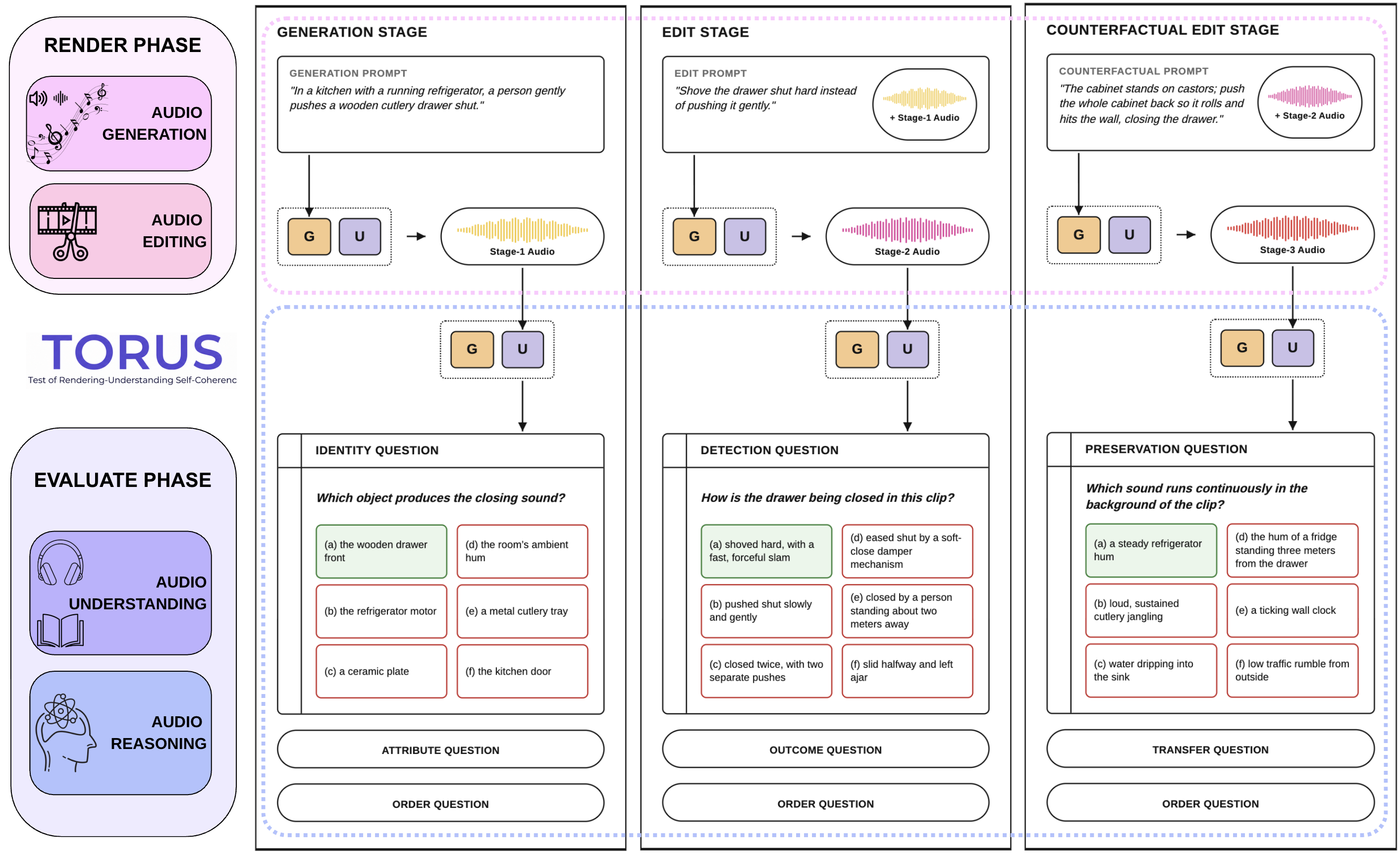}
\vspace{-10pt}
\caption{\textbf{Evaluation pipeline.} Each test comprises 3 stages (9 MCQs) s.t. each stage has a Render ($G$ generates/edits a clip) and Evaluate Phase ($U$ answers three MCQs about it).}
\label{fig:case}
\end{figure*}

\begin{figure*}[t]
\centering
\includegraphics[width=\linewidth]{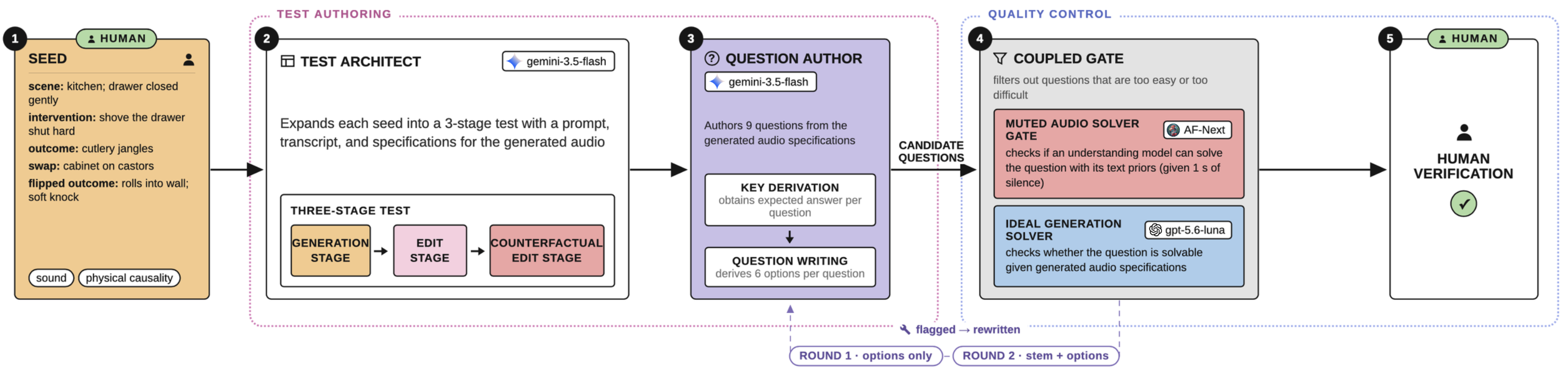}
\caption{\textbf{Test construction pipeline.} Human authoring $+$ verification at start \& end and LLM/LALM authoring $+$ gates.}
\label{fig:pipeline}
\end{figure*}


\vspace{-10pt}
\section{The \benchname{} Benchmark}
\label{sec:benchmark}

\benchname{} is designed to holistically evaluate unified
audio models and contains 117 \emph{tests} (39 tests per audio type), each containing 3 stages (Fig. \ref{fig:case}): generation, edit and counterfactual edit with 3 six-option multiple-choice questions each i.e. 9 MCQs per test (1{,}053 MCQs total). The benchmark spans 21 tasks (5 task families) and 8 audio domain settings (Fig.~\ref{fig:spider_and_stats}).
\vspace{-10pt}
\subsection{Task formulation and test structure}
\label{sec:taskform}

Each stage has a \emph{Render Phase}, in which $G$ generates/edits a clip, and an
\emph{Evaluate Phase}, in which $U$ answers the stage's three questions given only the clip and each question (Fig.~\ref{fig:case}). In the
\emph{generation stage}, $G$ renders a clip $a_{\mathrm{gen}}$ from a
generation prompt $p_{\mathrm{gen}}$. In the \emph{edit stage}, $G$ edits
$a_{\mathrm{gen}}$ under an edit prompt $p_{\mathrm{edit}}$, producing
$a_{\mathrm{edit}}$. In the \emph{counterfactual edit stage}, $G$ edits
$a_{\mathrm{edit}}$ under a counterfactual prompt $p_{\mathrm{cf}}$ (designed to change one attribute of the scene in a counterfactual manner without specifying the outcome), producing $a_{\mathrm{cf}}$. 
\vspace{-5pt}
\subsection{Data curation and test construction}
\label{sec:construction}

Tests are curated using the procedure and LLM/LALM models presented in Fig.~\ref{fig:pipeline}. A human-authored \emph{seed} specifies a scene, an intervention and
its acoustic outcome, the attribute that the counterfactual edit stage
changes, and the resulting counterfactual outcome. An LLM test architect expands the seed into the three stage prompts and corresponding ideal-audio specification: a text description ($+$transcript) of what a faithful clip for that prompt contains. An LLM question author selects three keys per stage from the ideal-audio specification and writes a six-option question for each key. A coupled gate then checks (and flags) every
question with two solvers: \emph{muted-audio solver}
(\texttt{Audio-Flamingo-Next}~\cite{ghosh2026afnext}) answers the question
given 1s silence (correct answer $\equiv$ the question is answerable from text priors) \& \emph{ideal-generation solver} answers the question given
the ideal-audio specification (wrong answer $\equiv$ question is ambiguous even with perfect clip). A flagged question returns
to the question author for at most 2 repair rounds (options only,
then question $+$ options). Human annotators review if test adequately
and reasonably evaluates models.

\vspace{-10pt}
\section{Experimental Setup}
\label{sec:setup}


\begin{table*}[t]
\centering
\footnotesize
\setlength{\tabcolsep}{1.6pt}
\renewcommand{\arraystretch}{0.96}
\newcommand{\allt}{\lsp\lso\lmu}
\renewcommand{\pms}[1]{{\tiny$\pm$#1}} 
\newlength{\frw}\settowidth{\frw}{\lopenai\textbf{\ugcf{}}$\uparrow$} 
\newlength{\fcw}\settowidth{\fcw}{\textbf{53.6}\pms{3.6}}            
\ifdim\fcw<\dimexpr0.5\frw-\tabcolsep\relax \setlength{\fcw}{\dimexpr0.5\frw-\tabcolsep\relax}\fi
\newcolumntype{F}{w{c}{\fcw}} 
\newlength{\syscolw}\settowidth{\syscolw}{\lnvidia Audex-30B (\lselfcap)} 
\resizebox{\textwidth}{!}{
\begin{tabular}{@{}l|cc|FF|cc|cc|cc|cc|cccccc@{}}
\toprule
& \multicolumn{2}{c|}{\lself\textbf{\ugco{}}$\uparrow$} &
\multicolumn{2}{c|}{\lopenai\textbf{\ugcf{}}$\uparrow$} &
\multicolumn{2}{c|}{\textbf{WER (\%)}$\downarrow$} &
\multicolumn{2}{c|}{\textbf{CM (\%)}$\uparrow$} &
\multicolumn{2}{c|}{\textbf{KL}$\downarrow$} &
\multicolumn{2}{c|}{\textbf{FAD}$\downarrow$} &
\multicolumn{6}{c}{\textbf{PQ}$\uparrow$} \\
\cmidrule(lr){2-3} \cmidrule(lr){4-5} \cmidrule(lr){6-7} \cmidrule(lr){8-9}
\cmidrule(lr){10-11} \cmidrule(lr){12-13} \cmidrule(l){14-19}
 &
\lta & \ltts & \lta & \ltts &
\lta & \ltts &
\lta & \ltts &
\lta & \ltts &
\lta & \ltts &
\multicolumn{4}{c}{\lta} & \multicolumn{2}{c}{\ltts} \\
\cmidrule(lr){2-2} \cmidrule(lr){3-3} \cmidrule(lr){4-4} \cmidrule(lr){5-5}
\cmidrule(lr){6-6} \cmidrule(lr){7-7} \cmidrule(lr){8-8} \cmidrule(lr){9-9}
\cmidrule(lr){10-10} \cmidrule(lr){11-11} \cmidrule(lr){12-12} \cmidrule(lr){13-13}
\cmidrule(lr){14-17} \cmidrule(l){18-19}
\multirow{-3}{*}[8pt]{\diagbox[height=36pt,width=\dimexpr\syscolw+\tabcolsep\relax,outerleftsep=0pt,innerleftsep=1pt,innerrightsep=1pt]{Models \& \\ Setting}{Metrics \& \\ Setting}} &
\multicolumn{2}{c|}{\allt} & \multicolumn{2}{c|}{\allt} &
\lsp & \lsp & \allt & \allt &
\allt & \allt & \allt & \allt &
\lso & \lmu & \lsp & \allt & \lsp & \allt \\
\midrule
\lcasc Cascaded (\lnative)
  & - & - & - & 52.1\pms{3.6}
  & - & \textbf{3.8}\pms{2.3}
  & 57.5\pms{7.1} & 57.5\pms{7.1}
  & ref. & ref. & ref. & ref.
  & 5.33 & 6.17 & \textbf{7.90} & \textbf{6.47}\pms{.20} & \textbf{7.90} & 6.47\pms{.20} \\
\midrule
\lnvidia Audex-30B (\lselfcap)
  & \textbf{48.5}\pms{3.2} & 52.2\pms{3.3} & 50.4\pms{3.5} & 51.1\pms{3.5}
  & 133.7\pms{10.7} & 135.8\pms{23.4}
  & 57.8\pms{6.8} & 59.0\pms{6.7}
  & 3.13\pms{.22} & 2.31\pms{.20} & 4.76\pms{.76} & 7.17\pms{1.24}
  & \textbf{6.46} & \textbf{6.96} & 5.74 & 6.39\pms{.14} & 7.44 & \textbf{6.95}\pms{.12} \\
\ltencent Audio-Omni (\lselfcap)
  & 45.5\pms{3.2} & \textbf{52.4}\pms{3.9} & 50.7\pms{3.4} & \textbf{53.6}\pms{3.6}
  & 123.3\pms{12.9} & 126.3\pms{13.3}
  & \textbf{65.0}\pms{6.4} & \textbf{72.1}\pms{6.7}
  & 2.76\pms{.25} & \textbf{1.65}\pms{.15} & 4.83\pms{.92} & \textbf{4.69}\pms{.67}
  & 6.33 & 6.78 & 5.40 & 6.17\pms{.15} & 6.34 & 6.41\pms{.15} \\
\ltencent Audio-Omni (\lnative)
  & 45.1\pms{3.2} & - & \textbf{51.5}\pms{3.3} & -
  & 126.6\pms{14.5} & -
  & 61.0\pms{7.3} & -
  & \textbf{2.68}\pms{.28} & - & \textbf{4.58}\pms{.67} & -
  & 6.25 & 6.85 & 5.20 & 6.10\pms{.19} & - & - \\
\lnvidia Audex-2B (\lselfcap)
  & 38.2\pms{3.1} & 41.7\pms{3.2} & 51.3\pms{3.8} & 51.8\pms{3.7}
  & 132.2\pms{15.0} & 140.3\pms{31.6}
  & 58.7\pms{6.7} & 59.5\pms{6.8}
  & 3.31\pms{.22} & 2.56\pms{.23} & 5.53\pms{.71} & 7.47\pms{1.27}
  & 6.45 & 6.72 & 5.34 & 6.17\pms{.15} & 7.34 & 6.84\pms{.12} \\
\lnone UniAudio-2 (\lselfcap)
  & 25.3\pms{2.7} & 27.7\pms{3.0} & 46.5\pms{3.3} & 48.0\pms{3.4}
  & 138.3\pms{14.7} & 67.7\pms{5.0}
  & 41.6\pms{6.6} & 39.6\pms{6.4}
  & 3.36\pms{.23} & 2.14\pms{.23} & 6.45\pms{.99} & 6.51\pms{1.07}
  & 6.01 & 6.63 & 5.83 & 6.16\pms{.13} & 6.99 & 6.54\pms{.13} \\
\lai Unified-IO 2 (\lselfcap)
  & 21.7\pms{2.3} & - & 48.6\pms{3.6} & 48.6\pms{3.6}
  & \textbf{99.0}\pms{2.7} & -
  & 29.9\pms{6.0} & 29.9\pms{6.0}
  & 3.68\pms{.34} & 3.68\pms{.34} & 4.78\pms{.73} & 4.78\pms{.73}
  & 5.67 & 5.85 & 5.40 & 5.64\pms{.11} & - & - \\
\bottomrule
\end{tabular}}
\vspace{-10pt}
\caption{\textbf{Performance of models on \benchname{}.} \lself\ugco{} and
\lopenai\ugcf{} show that models struggle to understand their own generations. We report native (\lnative) self-caption chain (\lselfcap) editing settings. $\pm$: half-width 95\%ile bootstrap CI over tests (10{,}000 resamples). \ltts  $\equiv$ Speech tests in TTS mode. Row 3 shows audio type: speech (\lsp), sound (\lso), music (\lmu) or all. \textbf{Bold} is best.}
\label{tab:main}
\end{table*}

\textbf{Models.}\label{sec:goldsetup}\label{sec:chain} We evaluate 5 open unified audio models that both generate and understand
speech, sound and music: \ltencent Audio-Omni~\cite{tian2026audioomni},
\lnvidia Audex-30B and Audex-2B~\cite{kong2026audex},
UniAudio-2~\cite{uniaudio2_2026} and \lai Unified-IO~2~\cite{lu2024unifiedio2}.
We exclude Qwen3-Omni~\cite{xu2025qwen3omni}, AuK~\cite{ma2026auk} \& Ming-Lite-Omni~\cite{gong2025mingomni} (generate only speech) and MuMu-LLaMA~\cite{liu2024mumullamamultimodalmusicunderstanding} (generates music only). Audio-Omni is the only model that supports native editing (\lnative) in the setting \benchname{} is designed for. Due to limited support (\lnative), we propose the self-caption editing chain (\lselfcap) setting. At editing stages (2 \& 3), $U$ captions the previous stage's clip. Given the edit prompt and caption, $U$ writes a derived generation prompt which describes the resulting audio (and transcript). $G$ then generates a new clip from this derived generation prompt. Since the chain carries the scene through text, it may induce errors and so editing has been discussed only briefly. We thus evaluate \ltencent Audio-Omni in both settings (\lnative \& \lselfcap) and all other models using the self-caption editing chain. We also recognize that some models expose a TTS mode especially for text-to-speech generation and report results in both settings wherever applicable (\lai Unified-IO 2 and \ltencent Audio-Omni (\lnative) don't have a separate TTS mode). The
\emph{Cascaded Baseline} combines specialized models, routed by audio type: TangoFlux~\cite{hung2024tangoflux} generates sound, ACE-Step~1.5~\cite{gong2026acestep15} generates music, Step-Audio-EditX~\cite{yan2025stepaudioeditx} generates and edits speech, MMEdit~\cite{tao2025mmedit} edits sound and music, and \gptaudio{} serves as $U$; and estimates how coherent current systems are (not as an upper-bound). \lopenai \texttt{gpt-audio} was selected by agreement with the authors’ own answers on a 40-question sample (sampled uniformly at random across stages and audio types):  \lopenai \texttt{gpt-audio} matches human on 75.0\% while \lgemini\texttt{gemini-3.1-pro} matches on only (47.5\%). All generation/editing is run at authors' default settings and understanding inference run with greedy decoding on a single NVIDIA H100 (fixed seed).\\
\noindent\textbf{Evaluation.}
\label{sec:torus_metrics}
We measure understanding-generation coherence (UGC) as: \ugco{}, \% of total questions answered correctly by the system's own $U$ and \ugcf{}, \% questions answered correctly by a frontier understanding model (here, \gptaudio{}). Lack of UGC can be attributed to multiple factors: faulty generation/editing, weak understanding, etc. This decomposition helps us isolate the performance of both heads and estimate how well a stronger understanding model $U$ would perform given the same audio. We also measure clips directly: WER on Whisper large-v3~\cite{radford2023whisper} transcripts of speech clips v/s authored transcript; KL (PaSST~\cite{koutini2022passt} on AudioSet class); FAD, Fr\'echet audio (VGGish~\cite{hershey2017vggish}) both reported against Cascaded Baseline for uniformity; production quality (PQ, 1 to 10) from
Audiobox-Aesthetics~\cite{tjandra2025audiobox}; and the correct-modality rate (CM), \% clips whose top CLAP matches ground-truth (speech, sound, music and noise). 

\vspace{-10pt}
\section{Results and Discussion}
\vspace{-10pt}
\label{sec:results}
\begin{figure}[h]
    \centering
    \includegraphics[width=\linewidth]{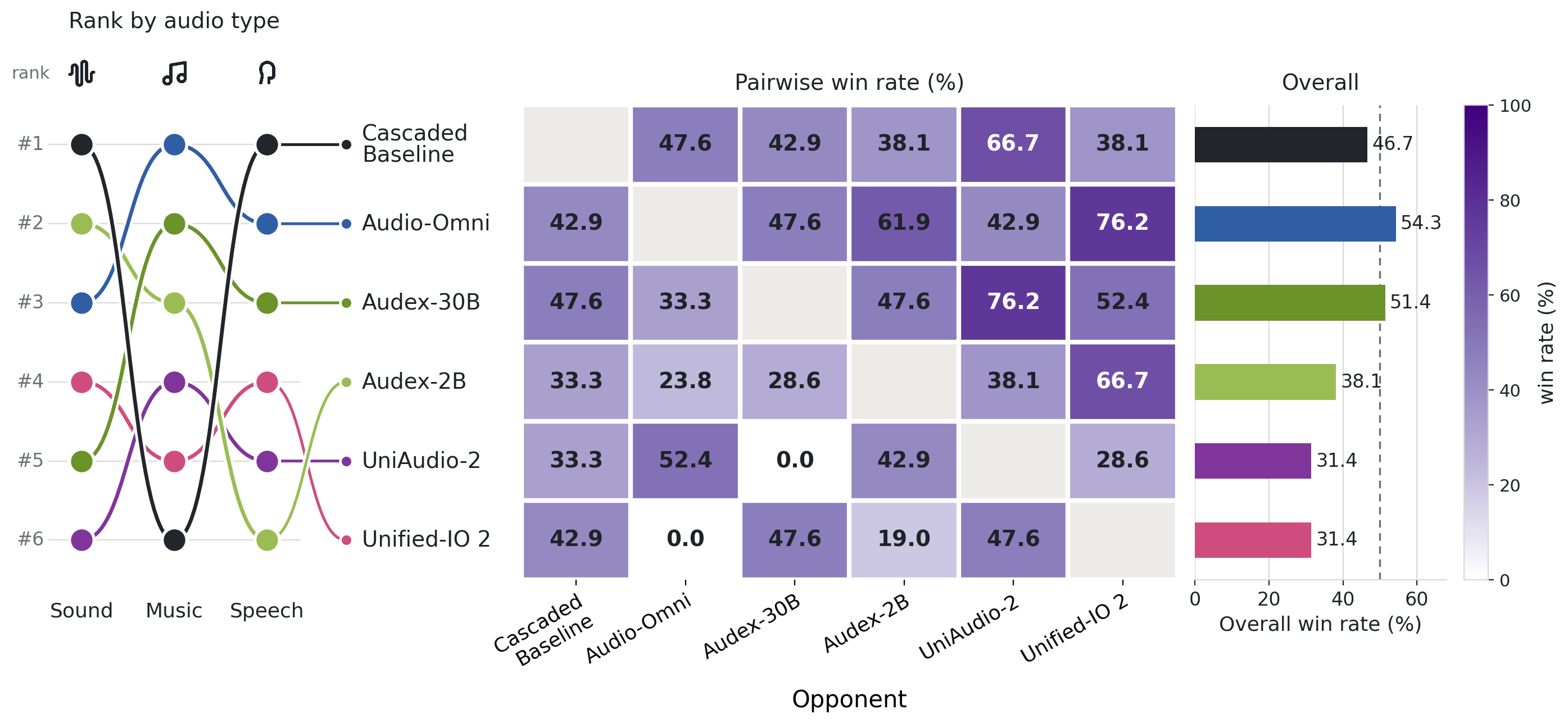}
    \vspace{-20pt}
    \caption{\textbf{Human A/B testing study} on model generations (7 raters, 315 judgments, Fleiss $\kappa$ = 0.377, Krippendorff $\alpha$ = 0.303) with audio-type wise ranks.}
    \label{fig:humaneval}
\end{figure}
\paragraph{Can unified audio models understand generated/edited audio?}\label{sec:lack_of_coh} We run all baselines in TTA \& TTS mode (wherever applicable) for speech and in TTA mode for music and sound tests, and \ltencent Audio-Omni in native and self-caption chain editing modes. We see that all evaluated unified audio models perform $\leq$ 52.4\% (chance 16.7\%) exhibiting limited understanding-generation self-coherence. We further see that performance of all baselines drops by up to 15.0\% (\ltencent Audio-Omni) at max  between the generation (1) \& editing (2\& 3) stages showing clear need for improvement in audio editing capabilities. However, since most baselines are run in a self-caption edit settings, we do not investigate this further and hope to see more eligible baselines developed.  We report \ugcf to understand how well frontier understanding models can understand same audio. While \ugcf > \ugco{} for most models, we see that \ugcf{} is always $\leq$ 53.6\% implying audio generation (across audio types) all evaluated systems leave room for improvement. \lnvidia Audex-30B scores \textbf{best} in \ugco-TTA (48.5\%) while \ltencent Audio-Omni scores \textbf{best} in \ugco-TTS, \ugcf-TTA and \ugcf-TTS (53.6\%) even higher than the 52.1\% Cascaded Baseline. This helps strengthen our hypothesis that unified models have the potential to show higher coherence than cascaded specialized models due to shared representations. However, given the current state of unified models and their performance, we cannot concretely confirm this hypothesis. On \textbf{audio fidelity} set-based metrics, we see that \ltencent Audio-Omni achieves lowest KL and FAD. Since these are calculated against the Cascaded Baseline, we resort to PQ and subjective human evaluation. We find that unified audio models \textbf{struggle most with speech tests} (in TTA and TTS modes) and trail cascaded specialists (PQ: 7.90 v 7.44 (\lnvidia Audex-30B \ltts)/5.83 (UniAudio-2 \lta)). On the other hand, most unified models outperform the Cascaded Baseline on music and sound PQ. We perform a \textbf{human A/B testing study} as presented in Fig. \ref{fig:humaneval} and find that raters prefer \ltencent Audio-Omni (\lnative) most (54.3\% Avg. win-rate) more than Audex-30B (51.4\%) and the Cascaded Baseline (46.7\%). \\
\noindent\textbf{Unified model performance ($U$ v $G$) bottlenecks}. On \textbf{speech tests}, models score 55.0\%, 64.1\%, 59.5\%, 48.7\%, 36.2\%, 23.9\% \ugco-TTS (same order as Fig. \ref{fig:humaneval}) hinting that generation is the weaker head and models have stronger speech understanding. Raters also rank the Cascaded Baseline first on speech, and T2A speech PQ stays at or below 5.83, against 7.90. On \textbf{sound tests}, the Cascaded Baseline scores 53.3\% while every unified
model scores lower (41.9, 48.1, 36.8, 22.8 and 20.8\%). The evidence points to
understanding as the weaker head: unified models exceed the baseline's sound
PQ, UGC $\leq$ 48.1\% even when raters rank Audex-2B, Audio-Omni and Unified-IO~2 high. Since  \ugco{} does not follow the rater ranking (even though Cascaded baseline performs best in both), we do not completely discount the generation head and encourage further research. On \textbf{music tests}, the best unified models exceed the Cascaded Baseline
(51.3 and 49.0\% against 48.1\%, \ugco), while the rest score lower (39.6, 24.2 and
20.5\%). The evidence points to generation as the stronger head: raters clearly prefer unified models and similar trends are seen in PQ as well, implying the bottleneck may lie in music understanding. Overall, we see understanding and generation performance varies significantly across audio types \& models implying audio generation and understanding both need to improve to achieve audio coherence (UGC).



\vspace{-10pt}

\section{Conclusion}
\label{sec:conclusion}

We presented \benchname{}, the 1\textsuperscript{st} closed-loop understanding-generation coherence evaluation benchmark for unified audio models curated using our human-verified framework. Our experiments show best unified models \ltencent Audio-Omni and \lnvidia Audex-30B achieve only 52.4\%/52.2\% self-coherence \ugco{} and  53.6\%/51.1\%  frontier \ugcf{}  on 1{,}053 \benchname{} questions exposing a large gap in how strong audio understanding and generation models (both specialized and unified) are. We further find that performance bottlenecks vary for understanding and generation across models and audio types and investigate each through objective metrics and human evaluation. We hope this work draws attention to the importance of self-coherence in unified models and leads to the emergence of more capable unified models.

\bibliographystyle{IEEEbib}
\bibliography{references}
\end{document}